\documentclass[aps,pra,twocolumn,amsmath,amssymb,superscriptaddress,10pt]{revtex4-1}
\usepackage{amsfonts}
\usepackage{amsmath}
\usepackage{amssymb}

\usepackage{graphicx}
\usepackage[T1]{fontenc}
\usepackage[utf8]{inputenc}
\usepackage{hyperref}
\usepackage[dvipsnames]{xcolor}
\usepackage{color}
\usepackage{bm}
\usepackage{comment}
\usepackage{mathtools}
\usepackage[section]{placeins}
\usepackage[capitalize]{cleveref}
\usepackage{footnote}
\usepackage{etoolbox}

\usepackage{pgfplots} 
\usetikzlibrary{plotmarks}
\pgfplotsset{compat=newest} 
\pgfplotsset{plot coordinates/math parser=false}

\crefrangelabelformat{equation}{(#3#1#4--#5\crefstripprefix{#1}{#2}#6)}

\begin{document}
\title{\texorpdfstring{LiHoF$_4$}{LiHoF4}: Cuboidal Demagnetizing Factor in an Ising Ferromagnet}

\author{M. Twengstr{\"o}m} 
\affiliation{Department of Physics, Royal Institute of Technology, SE-106 91 Stockholm, Sweden}
\author{L. Bovo} 
\affiliation{London Centre for Nanotechnology and Department of Physics and Astronomy, University College London, 17-19 Gordon Street, London, WC1H OAH, UK}
\affiliation{Department of Innovation and Enterprise, University College London, 90 Tottenham Court Rd, Fitzrovia, London W1T 4TJ, UK}
\author{O. A. Petrenko}
\affiliation{University of Warwick, Department of Physics, Coventry, CV4 7AL, UK}
\author{S. T. Bramwell}
\affiliation{London Centre for Nanotechnology and Department of Physics and Astronomy, University College London, 17-19 Gordon Street, London, WC1H OAH, UK}
\author{P. Henelius} 
\affiliation{Department of Physics, Royal Institute of Technology, SE-106 91 Stockholm, Sweden}
\affiliation{Faculty of Science and Engineering,  \r{A}bo Akademi University, \r{A}bo, Finland}

\begin{abstract}
The demagnetizing factor has an important effect on the physics of ferromagnets. For cuboidal samples it depends on susceptibility and the historic problem of determining this function continues to generate theoretical and experimental challenges. To test a recent theory, we measure the magnetic susceptibility of the Ising dipolar ferromagnet LiHoF$_4$, using samples of varying aspect ratio, and we reconsider the demagnetizing transformation necessary to obtain the intrinsic material susceptibility. Our experimental results confirm that the microscopic details of the material significantly affect the transformation, as predicted. In particular, we find that the uniaxial Ising spins require a demagnetizing transformation that differs from the one needed for Heisenberg spins and that use of the wrong demagnetizing transformation would result in unacceptably large errors in the measured physical properties of the system. Our results further shed light on the origin of the mysterious `flat' susceptibility of ordered ferromagnets by demonstrating that the intrinsic susceptibility of the ordered ferromagnetic phase is infinite, regardless of sample shape. 
\end{abstract}

\maketitle
\section{Introduction}
The demagnetizing energy of a magnetized sample presents several intriguing aspects. It plays a crucial role in the analysis of magnetic susceptibility~\cite{stoner45, osborn45}, realizes a laboratory example of long range interactions~\cite{prm_tweng_2017}, and even mediates some exotic physics {--}  for example, the complex nonlinear response and pattern formation in the intermediate state of type-I superconductors~\cite{Prozorov}. In view of the pioneering work of Poisson, Maxwell and others, the `demagnetizing effect' may at first sight appear to be a solved problem that belongs to the textbooks, but a closer appraisal of the literature reveals that it remains, to this day, a rather rich source of mathematical and practical challenges~\cite{chen02,chen05, beleggia, difratta16}. As far as magnetic materials are concerned, demagnetizing effects and corrections are particularly important in the discussion of several phenomena that are dominated by long range interactions, including, for example, magnetic monopole excitations in the spin ices~\cite{Castelnovo}, topological skyrmionic spin textures~\cite{Nagaosa} and spintronic applications of antiferromagnets~\cite{Jungwirth}. Even the problem of how to calculate the demagnetizing factor for shapes beyond ellipsoids is far from being solved in any general sense: many years of investigation have yielded some particularly elegant results~\cite{rowlands, rhodes} and ongoing work has revealed new surprises. In a recent study~\cite{prm_tweng_2017} we noted an unexpected dependence of the demagnetizing factor on the microscopic aspects of the material for rectangular prismatic samples, in contrast to the long-held expectation that only shape and macroscopic susceptibility should be relevant~\cite{chen02}. In this paper we elucidate this effect in detail with respect to a real model system {--} the Ising-like dipolar ferromagnet LiHoF$_4$. 

We therefore focus on a fundamental property of magnetism, namely the response to a small applied magnetic field, the uniform static magnetic susceptibility, $\chi=\lim_{H\to 0}\partial M/\partial H$ where $H$ is the internal magnetic field. The susceptibility is a fundamental thermodynamic characteristic of a magnetic system, reflecting its microscopic nature and magnetic state. With very careful measurement, it can reveal surprising properties, as exemplified in our recent detection of 'special temperatures' in frustrated magnets~\cite{Bovo18}. Compared to more local probes of spin correlations such as neutron scattering or muon relaxation, bulk susceptibility measurements offer the advantage of relative experimental simplicity and precise control of the experimental environment. However, there are still important aspects that one must consider in order to obtain a truly accurate measurement of the intrinsic material susceptibility. Many materials of recent interest contain high-moment rare-earth ions leading to a high susceptibility ($\chi\ge1$) and consequent strong demagnetizing effects. The state of the system can then only be defined and determined after particularly careful corrections for such effects. 

In detail, it is well-known that the internal magnetization and magnetic field of a paramagnetic ellipsoid exposed to an external magnetic field are uniform within the sample. The demagnetizing field $H_d$ is given by $H_d=-NM$, where $M$ is the magnetization, and $N$ the demagnetizing factor (more generally this is a tensor relationship $H_d^\alpha=-N^{\alpha \beta}M_\beta$). Remarkably, $N$, as defined in one direction, depends solely on the geometry of the sample and is independent of any underlying material properties. Due to the nature of the long-range dipolar interaction, the internal fields become non-uniform for non-ellipsoids, and the calculation of the demagnetizing factors a much more complex task. However, at an early point it was realized that it is possible to define a demagnetizing factor for cuboids that depends not only on the sample geometry, but also on the intrinsic susceptibility $\chi_{\textup{int}}(T)$, which leads to a temperature dependence of $N$~\cite{wuhr23,stab35}. Another avenue of research focused on the approximation of uniform magnetization, from which useful results were derived~\cite{rhodes, joseph}. Interestingly, the temperature dependent $N$ for cuboids has not been applied much in practice, although many experiments are performed on cuboids. 
This is despite the fact that the $\chi_{\textup{int}}${--}dependence of $N$ was calculated~\cite{chen02,chen05} nearly twenty years ago, using a finite-element method to solve the field equations. We and co-workers more recently introduced an alternative, iterative microscopic method, along with a brute force Monte Carlo calculation~\cite{prm_tweng_2017}, and the predicted $\chi_{\textup{int}}${--}dependence of $N$ was also supported by direct measurements on cuboids of the spin ice material Dy$_2$Ti$_2$O$_7$. However, as already mentioned, in addition to the $\chi_{\textup{int}}${--}dependence of $N$, Ref.~\cite{prm_tweng_2017} further discovered a dependence on the microscopic symmetry of the spin. For example, an isotropic Heisenberg spin, or isotropic multiaxial Ising material such as Dy$_2$Ti$_2$O$_7$, features a different $N$ from a uniaxial Ising material. In this study we test this theory using cuboids of the uniaxial Ising material LiHoF$_4$, and find that the experimentally determined $N$ matches the theory very well.

\section{Experimental Method}

\begin{figure}[!htb]
    \centering
    \resizebox{3cm}{!}{\includegraphics{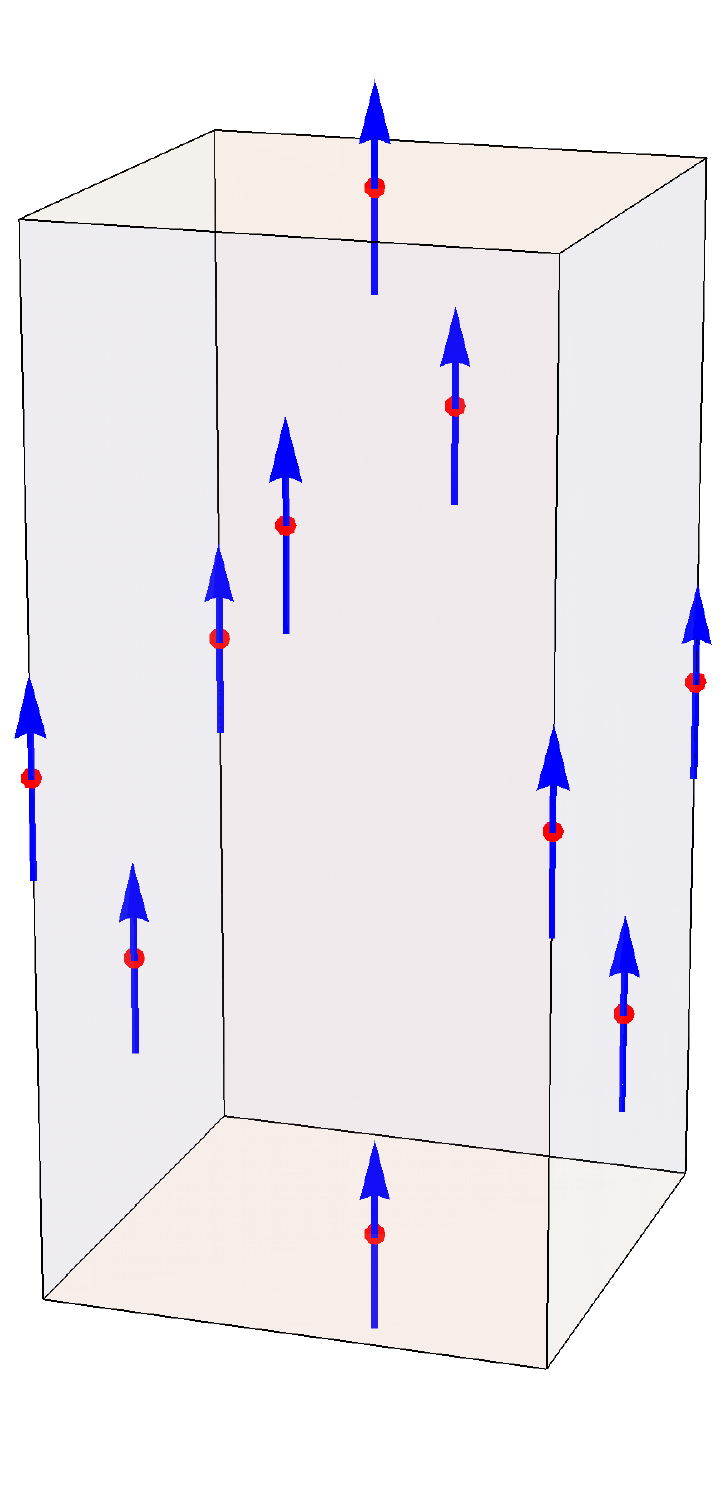}}
    \caption{The conventional unit cell of lithium holmium tetrafluoride (LiHoF$_4$). The red dots represent the holmium ion positions and the blue arrows indicate the Ising-like spins of LiHoF$_4$, when fully magnetized along its principal ($c$) axis.}
    \label{cell_label}
\end{figure}

LiHoF$_4$ (see~\cref{cell_label}) is an insulating rare-earth dipolar ferromagnet~\cite{ging11}. Due to the low-lying orbitals of the magnetic Ho$^{3+}$ ions, the dipolar interaction is stronger than the exchange interaction and the material orders magnetically at a relatively low critical temperature of $T_{\textup{c}}\approx 1.53$ K~\cite{menn84}. Significant crystal fields lead to a strong uniaxial Ising anisotropy with the Ising direction aligned with the principal axis of the tetragonal unit cell containing  four Ho$^{3+}$ ions, with a magnetic moment of about 7~$\mu_{\textup{B}}$ per ion. Due to the relative simplicity of the effective model~\cite{chak04}, the possibility to dilute the material with non-magnetic Y$^{3+}$ ions~\cite{reich87}, and its sensitivity to applied transverse magnetic fields, this compound has been used in numerous studies on classical and quantum phase transitions~\cite{griffin80, bitko96} and slow magnetic dynamics~\cite{bilt12}.

The crucial aspect of LiHoF$_4$ for this study is its uniaxial Ising symmetry, which distinguishes it from the isotropic, multiaxial Ising spin ice material  Dy$_2$Ti$_2$O$_7$ used in our previous study~\cite{prm_tweng_2017}. 
In addition, we benefited
from the commercial availability of LiHoF$_4$ aligned single crystals
cut to a range of sample shapes and aspect ratios that
would have been challenging to realize in a laboratory (or in-house) given the more brittle nature of this material.
In this investigation we  consider a spherical, cubic, long and needle sample of LiHoF$_4$, with dimensions given in~\cref{tab:shape}. The cuboidal crystals were grown, cut, aligned and polished by Altechna Co. Ltd. and we checked their alignment and crystal quality by X-ray Laue diffraction. Note that that the `cube' was not perfectly cubic, a difference that was accounted for in the analysis. The spherical crystal was derived from a sample supplied by Altechna, that was further hand cut as in Ref.~\cite{Bovo18}. It was accurately aligned along the $c$-axis  (the easy axis of magnetization) by applying a 7 T  magnetic field in a viscous liquid grease at room temperature and then cooling to solidify the grease. In general, the best estimate for the volume (used to determine the susceptibility) was calculated through the weight and the density, $\rho=5.72 \:\textup{g/cm}^3$. 

\begin{table}[!htb]
    \caption{Physical dimensions of samples used. Error bars are $\sim 0.03$ mm on dimensions, and $1$ in the last stated digit on weights. The last two columns contain the calculated demagnetizing factors in the limit of zero and infinite susceptibility for each sample.}
    \vspace{0.5cm}
    \centering
     \begin{tabular}{ | l | l | l | l |l|}
        \hline
        Shape & Dimensions [mm] & Weight [g] & $N(\chi= 0)$ &  $N(\chi=\infty)$ \\ \hline
        Sphere&{\O} $=3.8$ & 0.16415 & 1/3&1/3\\ \hline
        Cube & $4.08\times 3.87\times 4.09$ &0.36505   & 0.327 &0.274\\ \hline
        Long & $1.95\times 2.17 \times 8.10$  & 0.19235 & 0.110 &0.0751\\\hline
        Needle & $0.60\times 0.67\times 8.05$  &  0.01773 & 0.0363&0.0197\\\hline
    \end{tabular}
    \label{tab:shape}
\end{table}

The magnetic susceptibility was studied on different instruments in two temperature regimes, $ T\ge 1.8$ K and $ T\le 2$ K.

At $T \gtrsim 1.8$ K, the magnetic moment for each sample was measured as a function of temperature using a Quantum Design SQUID magnetometer, where the crystals were positioned in a cylindrical plastic tube to ensure a uniform magnetic environment. Measurements were performed in the RSO (Reciprocating Sample Option) operating mode to achieve improved sensitivity by eliminating low-frequency noise. For the cuboids we initially trusted the design specifications and used the crystal edges as reference for alignment in the magnetometer (our X-ray study later showed one crystal to be slightly misaligned, as discussed subsequently). By analogy with Ref.~\cite{Bovo}, different measurements were made: low field susceptibility and field-cooled (FC) versus zero-field-cooled (ZFC) susceptibility, with no significant differences observed. Also, magnetic field sweeps up to several hundred Oe at fixed temperature were performed in order to evaluate the susceptibility accurately, to confirm the linear approximation and estimate the absolute susceptibilities, following the method described in our previous work~\cite{Bovo}.

The magnetic moment at lower temperatures was measured using a different Quantum Design MPMS SQUID magnetometer equipped with an~\emph{i}Quantum $^{3}$He insert~\cite{iQuantum}. The applied fields were 50 and 100 Oe. 

Data between the high and low temperature regimes have been compared, in particular in the overlapping region $1.8 \le T \le 2$ K. Without further manipulation, the two sets of data are in very good agreement with variations of the order of $1 \%$. This variation can be attributed to several factors, including the uncertainty in the actual field value in each of the two instruments, (due in part to the presence of small frozen fields in the superconducting coils) and variations in precise sample positioning within the pickup coils. Here in the manuscript we only show data below $5$ K.

The field, measured in {\O}rsteds, and the magnetic moment $m$, measured in emu, were converted into SI units using
\begin{equation}
    \chi_{\textup{SI}}=\frac{4 \pi m[\textup{emu}]}{H[\textup{Oe}] V[\textup{cm}^3]}.
\end{equation}

\section{Theory of the demagnetizing factor}
The theory that we apply is given in detail in Ref.~\cite{prm_tweng_2017}, but it is useful to summarize some of its key aspects here. 

Textbook presentations of the demagnetizing factor emphasize how the homogeneity of the internal field and local magnetization for ellipsoids allows one to define a demagnetizing factor, $N$ for a specified crystal axis. This is a fixed number for any given ellipsoid: for example, the exact  demagnetizing transformation for a sphere ($N = 1/3$) is 
\begin{equation}
    \frac{1}{\chi_{\textup{int}}}=\frac{1}{\chi_{\textup{exp, sphere}}}-\frac{1}{3},
    \label{demag0}
\end{equation}
where $\chi_{\textup{exp, sphere}}  = \partial M/\partial H_{0}$ is the experimentally determined susceptibility and $H_0$ is the uniform applied magnetic field. 

This is typically contrasted with the case of non-ellipsoids, where neither internal field nor magnetization are uniform, with the consequence that a demagnetizing factor $N$ can no longer be defined as a unique, fixed number, in the way it can for ellipsoids. Nevertheless, for our purposes, it is most important to stress the fact that a temperature{--}dependent demagnetizing factor, $N(T)$ can still be precisely defined for any sample shape.  

To see this, consider an arbitrarily shaped sample, subject to the field $H_0$. The incremental magnetic work is $\mu_0 H_0 {\textup{d}} m$, where $m = M V$ is the magnetic moment of the sample of volume $V$, which uniquely defines the magnetization $M$.  We can then write
\begin{equation}
    \frac{1}{\chi_{\textup{int}}}=\frac{1}{\chi_{\textup{exp, ne}}}-N,
     \label{demag1}
\end{equation}
where $N(T)$ is the demagnetizing factor of the non-ellipsoidal (ne) sample, which corresponds to the standard `magnetometric' demagnetizing factor for simple shapes like cylinders or rectangular prisms. 

It can then be seen by eliminating $\chi_{\textup{int}}$ from~\cref{demag0,demag1}, that $N(T)$ is the quantity that precisely maps the temperature dependence of the magnetic moment of the non{--}ellipsoid  onto that of the sphere, or any arbitrary ellipsoid. Hence, a knowledge of $N(T)$ allows the measurement of $\chi_{\textup{int}}$ for any sample shape. In the fundamental investigation of magnetic materials, $\chi_{\textup{int}}$ is the quantity of interest as this can be calculated, in principle, from a knowledge of the spin Hamiltonian, or simulated numerically using periodic boundaries and Ewald methods. Here, the thermodynamic limit is taken in regards to the change of the surface term stemming from the fluctuations of the magnetic moments which produce a surface charge. The full derivation was made by de Leeuw~\emph{et al.} in Ref.~\cite{leeuw80a} using a semi-classical approach and a thorough microscopic derivation on the matter can be found in Ref.~\cite{perram87}.

For a system with inhomogeneous fields, it is therefore still possible to precisely define $N(T)$, without making explicit reference to the inhomogeneities. These do, however, continue to play a crucial role in determining the numerical value $N(T)$ at each temperature. For a given $\chi_{\textup{int}}$, our iterative method accounts for the inhomogeneity spin by spin and can be implemented on any given spin structure, or local spin symmetry (Ising, XY, Heisenberg), for system sizes up to about $\mathcal{N} = 10^6$ spins. We supplement it by direct, brute force, Monte Carlo simulations of the spin{--}Hamiltonian, and in both cases extrapolate to the thermodynamic limit, $\mathcal{N} \rightarrow \infty$. In Ref.~\cite{prm_tweng_2017} we and colleagues showed that the thermodynamic limit values of $N(T)$ are in excellent agreement for the two methods, even though the finite size corrections are rather different. 

We believe that these methods go beyond previous approaches in that the viewpoint is no longer mesoscopic (i.e. `micro'-magnetic in the common notation), but rather truly microscopic and spin{--}Hamiltonian based. Hence it is more appropriate for certain fundamental studies, such as that of spin ice~\cite{prm_tweng_2017} and LiHoF$_4$, studied here. This does not detract from the value of the micromagnetic approaches for many magnetic problems and we have demonstrated complete agreement between our approach and that of Chen~\emph{et al.}~\cite{chen02,chen05} in the case of cubic spin ice. 

Indeed the works of Chen et al. have revealed many important features of the problem: notably, for cylinders (and we can expect the same for cuboids) as $\chi\rightarrow 0$ the demagnetizing field is non-uniform but the magnetization is uniform~\cite{chen91}. Hence, in that limit, using the results of Ref.~\cite{rhodes}, the cube has the same demagnetizing factor as a sphere, $N=1/3$. In the opposite limit, $\chi\rightarrow \infty$, the roles are reversed and the demagnetizing field is uniform (on some mesoscopic scale) but the magnetization is non-uniform~\cite{chen91}. So, for a cube, $N$ takes a different limiting value, $N \approx 0.27$. Our work identifies new aspects of the behavior of the function $N(\chi)$ between these two limits.

In our method, we first determine the function $N(\chi_{\textup{int}})$, which depends only on the definition of the spin degrees of freedom and their dipole-dipole interactions, and importantly, is independent of exchange terms in the spin Hamiltonian. Then, by substituting $\chi_{\textup{int}}(T)$ into $N(\chi_{\textup{int}})$, the new function 
$N(T)$ is determined and it is at this point that the full details of the spin Hamiltonian enter into the problem. Hence all relevant terms in the spin Hamiltonian affect $N(T)$ but only dipolar terms affect $N(\chi_{\textup{int}})$.

The effect of anisotropy terms in the spin Hamiltonian is rather subtle.  They will in general affect the bulk susceptibility tensor and through that, the demagnetizing tensor and demagnetizing factor $N(\chi_{\textup{int}})$, in a way that is compatible with the crystal symmetry. For example, spin ice has local Ising spins but its space symmetry is cubic. Hence, the local Ising terms are not manifest in the function $N(\chi_{\textup{int}})$, which is the same as that of other isotropic systems. However, they do strongly affect the temperature dependence of $\chi_{\textup{int}}(T)$ and through that, the function $N(T)$. 

For the uniaxial spin system studied in this paper, and for a given sample shape, $N(\chi_{\textup{int}})$ will be a different function to that of cubic spin ice because the susceptibility tensor has different symmetry, although it will coincide in the limit $\chi\rightarrow 0$, where the magnetization becomes homogeneous and also, it seems~\cite{prm_tweng_2017}, in the limit $\chi\rightarrow \infty$,  where the demagnetizing field becomes homogeneous (see above). It is in this rather subtle way, where $N(\chi)$ and $N(T)$ are both affected, but to different degrees, that microscopic effects {--} and in particular the effects of local spin symmetry {--} may be revealed in the behavior of the demagnetizing factor for non{--}ellipsoidal samples, such as the cuboids studied here.

\section{Results}
\begin{figure}[!htb]
    \resizebox{\hsize}{!}{\includegraphics{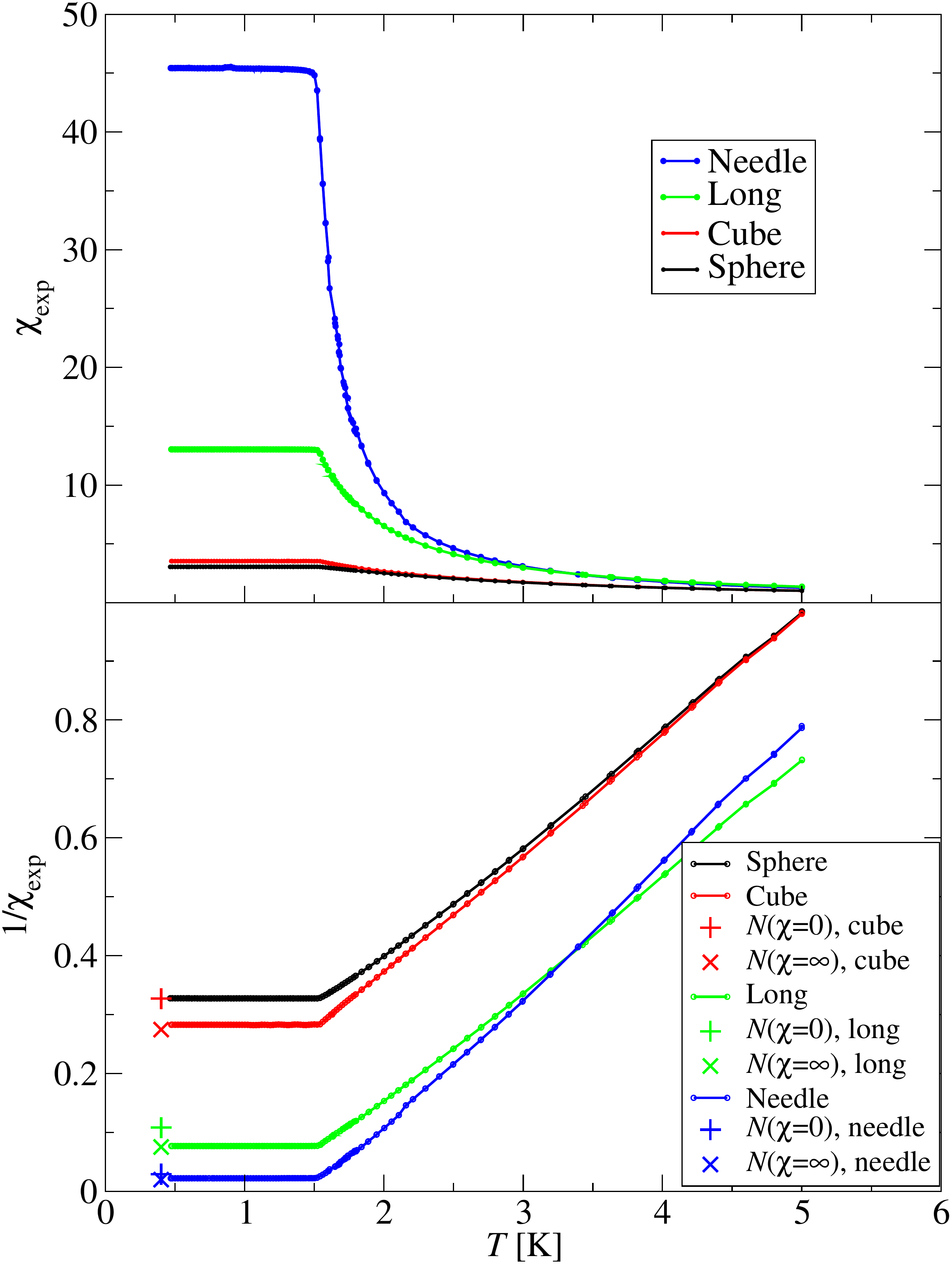}}
    \caption{The experimentally measured susceptibility (upper panel) and inverse susceptibility (lower panel), as a function of temperature, $T$, for the differently shaped samples of LiHoF$_4$ listed in Table 1. The lower plot also shows $\chi\rightarrow 0$ and $\chi\rightarrow \infty$ values of $N(\chi)$, from the tables of Ref.~\cite{chen05}.}
    \label{sus_combo}
\end{figure}

In~\cref{sus_combo} (upper panel) we show the measured low{--}temperature susceptibility of the different samples, and note that the shape{--}dependence dominates the  susceptibility. 

The inverse susceptibility, shown in~\cref{sus_combo} (lower panel), is very reminiscent of Fig.~2 in the detailed early study of Cooke~\emph{et al.}~[\onlinecite{jphys_cooke_1975}]. The low{--}$T$ plateau in the inverse susceptibility below the critical temperature of the material is expected to occur at the value of the demagnetizing factor $N$ for the sample (see Discussion). The low{--}$T$ susceptibility thus provides direct experimental access to the demagnetizing factor for ferromagnets. The estimated $\chi\rightarrow 0$ and $\chi\rightarrow \infty$ values of $N$~\cite{chen02,chen05}, indicated respectively by a plus and a cross on the graph, are compared to the low{--}temperature plateau of the susceptibility for each sample. The $\chi\rightarrow \infty$ values are significantly closer to experiment, a first significant result that we return to below in the Discussion. 

For the spherical sample we would expect the plateau exactly at $\chi=1/N=3$, while it is a bit higher $\chi=3.037$, possibly due to a deviation from a perfectly spherical shape. The slope in the limit of high temperature should be identical for all samples since it is related to  the Curie constant of the material. The gradients in the 4{--}5 K interval are similar for the sphere and cube (0.199 and 0.203) respectively, but higher for the long sample and needle (0.220 and 0.213 respectively). This discrepancy is larger than what our theory can account for, and we suspect that it is due to a misalignment between the local Ising axis and the long side of the cuboids. Using a Laue camera, we estimate the misalignment to 5 and 6 degrees for the long and needle samples respectively, supporting this hypothesis.  

If the demagnetizing factor happened to be independent of temperature, as many studies assume, all the curves in~\cref{sus_combo} (lower) should simply be vertically shifted images of each other, as can be deduced from the classical demagnetizing transformation
\begin{equation}
    \frac{1}{\chi_{\textup{int}}}=\frac{1}{\chi_{\textup{exp}}}-N.
    \label{demag}
\end{equation}
However, in~\cref{sus_combo}, we clearly see that this is not true for the spherical and cubical samples, since the curves start to diverge at low temperature. These temperature{--}dependent deviations from the usual demagnetizing transformation are the main subject of this study.

Since the demagnetizing factor for the sphere is independent of the temperature, we can find the intrinsic susceptibility of the material using~\cref{demag}  with $N=1/3.037=0.3293$ in the present case. Using the intrinsic susceptibility we then obtain the temperature dependent demagnetizing factor $N$ for the other samples using~\cref{demag}.

\begin{figure}[!htb]
    \resizebox{\hsize}{!}{\includegraphics{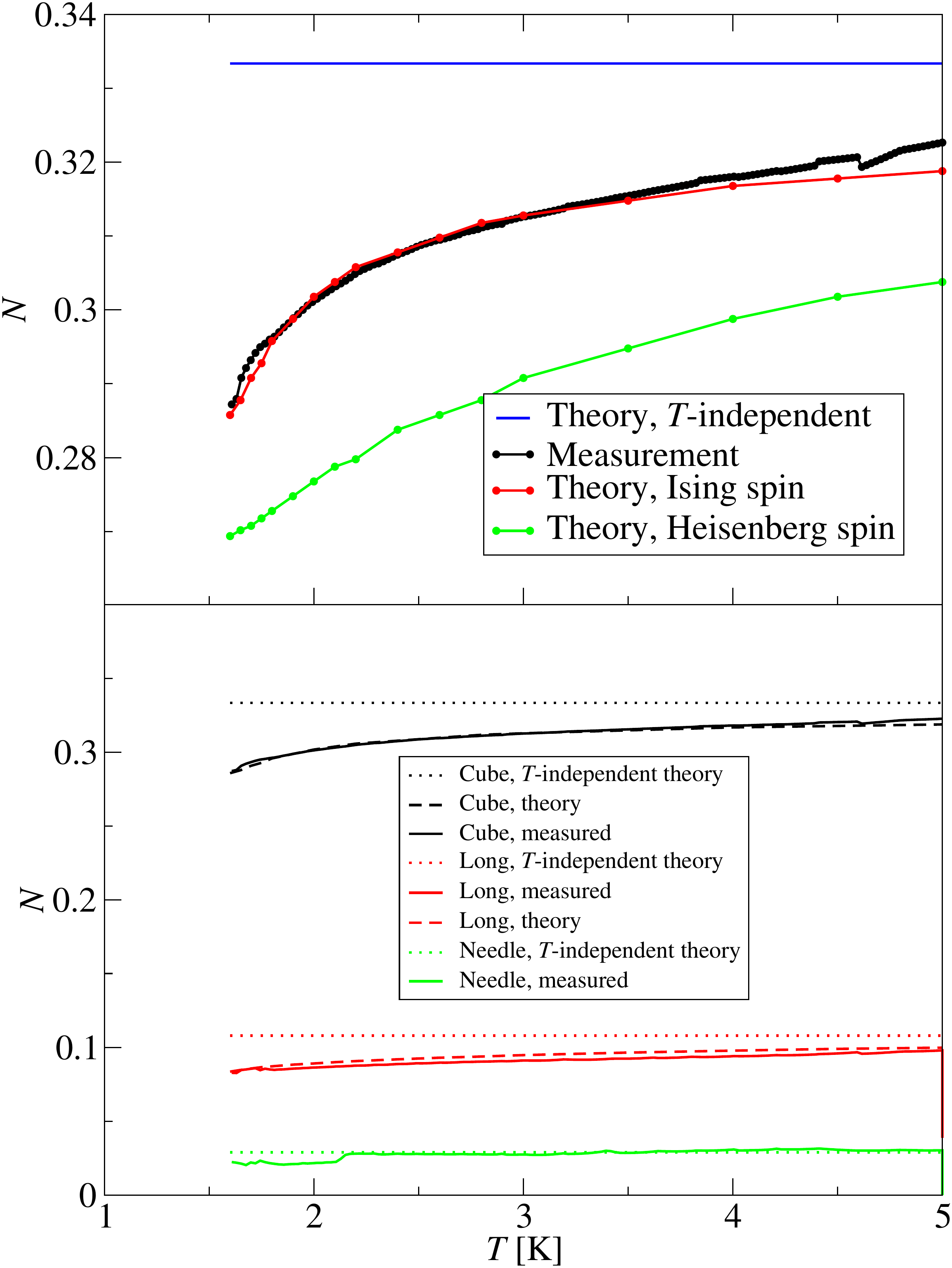}}
    \caption{Upper panel: Temperature dependence of the experimentally derived demagnetizing factor $N(T)$ for the approximate cube of LiHoF$_4$ (black points) compared with theoretical predictions (blue, red, green). The prediction for Ising spins accounts for the experimental data very accurately, while other predictions fail. Lower panel: Temperature dependence of the demagnetizing factor (both experimental and theoretical) for the differently shaped samples of LiHoF$_4$ listed in Table 1. Our theory (referred to as `Theory, Ising spin' on the figure) accurately accounts for all experimental observations.}
    \label{n_combo}
\end{figure}

The main result of this study is shown in~\cref{n_combo} (upper panel), where we see the experimentally determined $N$ as a function of $T$ for the cube along with our theoretical predictions for a uniaxial Ising material such as LiHoF$_4$. The details of the calculation are given in Ref.~\cite{prm_tweng_2017}. In addition, we show the commonly assumed $T${--}independent value of $1/3$ (equal to $N(\chi \rightarrow 0)$ ), as well as the theoretical predictions for an isotropic material~\cite{chen02,chen05,prm_tweng_2017}, which both differ significantly from the true result. In~\cref{n_combo} (lower panel), $N(T)$ is shown for all the samples, along with the commonly used $T${--}independent theory, and our theoretical calculation, which takes the symmetry of the spin and $T${--}dependence into account. The agreement between experiment and our theory is very satisfactory in all cases, while the $T${--}independent ($\chi\rightarrow 0$) theory fails to describe experiment.

In order to determine $N(T)$ for the long sample and needle, we multiplied the susceptibilities with factors $1.085$ and $1.07$ respectively, to ensure that the slope of $\chi(T)$ approaches the same high{--}$T$ limit, which is a physical requirement. As noted above, we are not certain what the source of this deviation is, but strongly suspect the verified misalignment of the local magnetization axis with respect to the main axis of the sample.

\begin{figure}[!htb]
    \resizebox{\hsize}{!}{\includegraphics{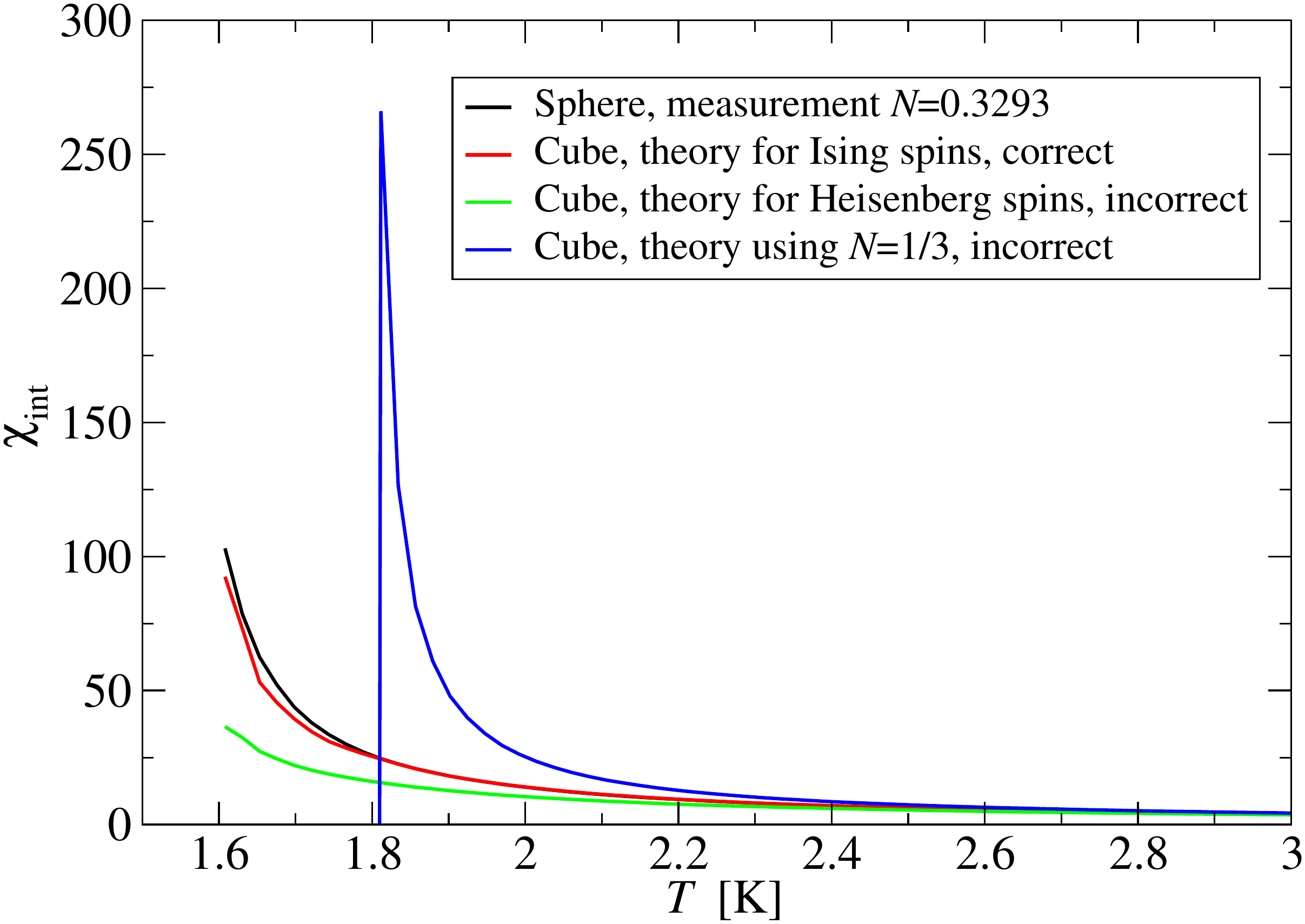}}
    \caption{Experimentally derived intrinsic susceptibility of LiHoF$_4$ derived under different assumptions as to the behavior of the demagnetizing factor for cuboidal samples. The true value is determined from the susceptibility of the sphere using $N=0.3293$ (black) and this is compared with that derived from the cube using our theory for uniaxial Ising spins (red), the theory for Heisenberg spin (green) and the temperature independent commonly used value of $N=1/3$. It can be seen that use of an incorrect demagnetizing factor leads to unacceptable errors in the derived susceptibility.}
    \label{chi_int}
\end{figure}

To emphasize the importance of accurate demagnetizing transformations when the susceptibility is large, like it is for many rare{--}earth based magnets at low temperature, we show in~\cref{chi_int} the result of determining the intrinsic material susceptibility from the measurement on the cube. In black, we show the reference intrinsic susceptibility determined from the sphere, using $N=0.3293$. For comparison, we have transformed the measurement on the cube in three different ways: using the theory for uniaxial Ising spins (red), the theory for Heisenberg spin (green) and the frequently used approach of assuming the high{--}susceptibility value of $N=1/3$. As~\cref{chi_int} illustrates, the theory for uniaxial Ising spins reproduces the reference susceptibility very well, while the theory for Heisenberg spins underestimate the susceptibility considerably {--}  by more than 50 \% at the lowest temperature. This would cause unacceptable errors in say, the determination of a critical exponent for the transition. Similarly, the $T${--}independent transformation diverges at $\chi_\textup{exp}=3$, which would falsely indicate a critical temperature of $T_{\textup{c}}=1.8$ K, well above the actual $T_{\textup{c}}=1.53\textup{ K.}$

\section{Discussion and Conclusion}
In summary, the demagnetizing transformation for an ellipsoid involves a single number that depends only on sample shape. For a non-ellipsoidal sample the transformation is still well-defined, but becomes much more subtle. Taking a cubic sample as an example, in the small $\chi$ limit, the transformation is (surprisingly) the same as that of a sphere, $N = 1/3$ and in the large $\chi$ limit, it reaches a number that does not depend on any microscopic details of the spin Hamiltonian. Between these limits, as we have demonstrated by comparing experiment to microscopic theory,  $N(\chi)$ depends on the underlying microscopic symmetry of the magnetic moment of the material. In~\cref{n_combo}, it is confirmed that the measured demagnetizing factor $N$ for the nearly cubic sample of LiHoF$_4$ agrees well with our theory, which takes the symmetry of the spin and the $T${-}dependence into account, while the measurement differs from both the $T${-}dependent result for an isotropic material~\cite{chen02,chen05,prm_tweng_2017}, and the $T${-}independent, small-$\chi$ value. In~\cref{chi_int}, we see that this seemingly small difference in $N$ has a very significant effect on the final transformed intrinsic susceptibility of the material, the aim of most susceptibility measurements. 
 
It is clear that use of an ellipsoidal sample, where the temperature independent demagnetizing factor is known, is a robust way to accurately determine the intrinsic susceptibility of a material. We note that early measurements of critical exponents on LiHoF$_4$~\cite{beauvillain} and, for example, the rare earth dipolar magnets RCl$_3\cdot 6$H$_2$O (R = Dy, Er)~\cite{lagendijk} and dysprosium ethyl sulfate~\cite{frowein}, did indeed employ ellipsoidal samples, so there is no reason to doubt their conclusions. Earlier studies on non-ellipsoidal samples may contain systematic errors, but the results presented here show how accurate intrinsic susceptibilities can be estimated from magnetization data on cuboidal samples.

Our measurements at temperatures below $T_{\textup{c}}$ confirm that the magnetic moment of the cuboidal samples remains remarkably constant, at the value $m = H_0 V/N(\chi= \infty)$.  This may be simply derived by setting the internal field to zero in the equation:
\begin{equation}
H_{\textup{int}} = H_0 - N M.
\end{equation}
Indeed it is a well-known property of many (typically `soft') ferromagnets that was discussed theoretically many years ago~\cite{arrott,wok,griffiths}. It appears that the closest to an explanation of this experimental fact was that found by Wojtowicz and Rayl~\cite{wok}, who considered a highly idealized model of a toroidal sample, where a perpendicular ordering field competes with a curling mode within the plane of the toroid: in that case, a mean field treatment yielded the observed behavior. This is an interesting result as it suggests a topological origin to the experimental observation of constant moment. However, in the present case of a uniaxial magnet it is difficult to make the same argument as the low temperature state is not a curling mode, but rather, a complex domain state. In  early studies of several uniaxial systems including LiHoF$_4$ (see Ref.~\cite{jphys_cooke_1975} and references therein), it was proposed that domain wall movement is sufficiently free that domains move so that the average demagnetizing field exactly cancels the applied field. Certainly, if the response is confined to the movement of purely macroscopic objects (the domain walls), this would be associated with effectively zero entropy change per spin and hence athermal behavior. If the free energy is equated to the demagnetizing energy $E=(\mu_0V/2)N M^2$ then  $\chi_{\textup{exp}} = 1/N$, as observed. 

However, for non-ellipsoidal samples, this raises the question as to which demagnetizing factor to use in the calculation of the moment. Our results (\cref{sus_combo}, lower panel) show conclusively that it is indeed the $\chi=\infty$ value of $N(\chi)$, as calculated here and in Refs.~\cite{chen02,chen05}, rather than the usual $\chi = 0$ value. As predicted in Ref.~\cite{wasilewski}, the paramagnetic fluctuations seem to anticipate the low temperature domain structure.

This result strongly indicates that $\chi_{\textup{int}} = \infty$ for all $T\le T_{\textup{c}}$, regardless of sample shape. In turn, it raises a certain ambiguity as to what $\chi_{\textup{int}}$ represents in the ordered phase. If $\chi_{\textup{int}}$ is interpreted as the susceptibility for $N = 0$ then one has to admit that it should be finite below $T_{\textup{c}}$, on account of the broken symmetry of the ferromagnetic state. On the other hand if it is defined as in~\cref{demag1} to be a property of a sphere (say), then it can be infinite as a property of the spherical domain state. As the temperature is lowered below $T_{\textup{c}}$, the domain magnetization will increase, and the domain susceptibility will decrease, consistent with a dipolar ordering transition~\cite{beauvillain, griffin}, but the bulk intrinsic susceptibility will remain infinite. The infinite susceptibility, or `critical line', at all temperatures below $T_{\textup{c}}$ is reminiscent of a topologically ordered Kosterlitz-Thouless phase~\cite{KT}, or of a soft mode (infinite transverse susceptibility) in an ordered, continuously degenerate system. In view of the Wojtowicz-Rayl argument~\cite{wok}, where the ferromagnetic transition is accompanied by the appearance of a global topological defect (the winding mode of a torus), and the infinite susceptibility as $N\rightarrow 0$ does arise from a soft mode, such analogies are worth considering.

A detailed numerical study of domain patterns in LiHoF$_4$ revealed a preference for a structure of parallel (to $c$) sheets of alternately spin `up' and spin `down'~\cite{biltmo09}. The infinite susceptibility could then reflect the free motion of smooth or rough domain walls that restore symmetry (at least locally) between spin up and spin down ordered states. There is indeed a certain topological character to the phenomenon as domain walls can be classified as topological defects, and rough ones can even map microscopically to the Kosterlitz-Thouless phase (although the evidence is that the long range interaction suppresses roughness in LiHoF$_4$~\cite{Mias_Girvin_2005}). Further investigation of the topological origins of the flat susceptibility of ferromagnets would certainly be worthwhile.

It is finally worth emphasizing that the corrections to the demagnetizing transformation that we have identified are, of course, indicative of inhomogeneous fields within the sample. Such inhomogeneities are likely to be macroscopic, with details on length scales that are not much shorter than the sample dimensions~\cite{biltmo09}. Thus diffuse magnetic neutron scattering, that measures generalized susceptibilities on rather smaller scales, is not likely to be strongly affected by these corrections, but magnetic Bragg scattering will be strongly affected {--} a fact that will need to be accounted for in neutron scattering studies of ordered states. In general, from the perspective of magnetic moment measurement, field inhomogeneities represent a correction to be transformed away, but from a more general perspective, they are an interesting phenomenon in their own right and can be precisely analyzed, as we have illustrated in this paper.

\begin{acknowledgments}
 The simulations were performed on resources provided by the Swedish National Infrastructure for Computing (SNIC) at the Center for High Performance Computing (PDC) at the Royal Institute of Technology (KTH). We gratefully acknowledge the NVIDIA Corporation for the donation of GPU resources. M.T. was supported by Stiftelsen Olle Engkvist Byggmästare (187-0013) with support from Magnus Bergvalls Stiftelse (2018-02701). L.B. was supported by The Leverhulme Trust through the Early Career Fellowship programme (ECF2014-284).

The authors declare no competing financial interests.
\end{acknowledgments}

\end{document}